\documentclass[aps,11pt]{revtex4}
\usepackage[dvips]{graphicx}
\begin{document}

\title{Orbital and spin moment in CoO$^\ast$}
\author{R. J. Radwanski}
\homepage{http://www.css-physics.edu.pl} \email{sfradwan@cyf-kr.edu.pl}
\affiliation{Center for Solid State Physics, S$^{nt}$Filip 5, 31-150 Krakow, Poland,\\
@ Institute of Physics, Pedagogical University, 30-084 Krakow, Poland}
\author{Z. Ropka}
\affiliation{Center for Solid State Physics, S$^{nt}$Filip 5, 31-150 Krakow, Poland}
\begin{abstract}
The orbital and spin moment of the Co$^{2+}$ ion in CoO has been calculated within the quasi-atomic approach with
taking into account the intra-atomic spin-orbit coupling and crystal field interactions. The orbital moment of 1.38
$\mu _{B}$ amounts at 0 K, in the magnetically-ordered state, to more than 34\% of the total moment (4.02 $\mu _{B}$)
and yields the L/S ratio of 1.04, close to the recent experimental value. This paper has been motivated by a recent
paper in Phys. Rev. B \textbf{65} (2002) 125111, in which a presented theoretical approach was unable to calculate the
orbital moment.


\end{abstract}

\maketitle

\section{Introduction}

A consistent description of properties of CoO, in general of 3d-atom oxides, including monooxides, reconciling its
insulating state with the unfilled 3$d$
shell is still not reached as one can learnt from recent papers \cite%
{1,2,3,4,5,6,7,8}.

The aim of this short paper, \cite{9}, is to report the calculations of the magnetic moment of CoO, its spin and
orbital contribution. The direct motivation was a recent paper of Ref. \cite{8}, published in Phys. Rev. B \textbf{65}
(2002) 125111, in which a presented theoretical approach was unable to calculate the orbital moment. In our approach we
attribute the
moment of CoO to the Co$^{2+}$ ions. We have calculated the moment of the Co$%
^{2+}$ ion in the CoO$_{6}$ octahedral complex, its spin and orbital parts. The orbital moment at 0 K as large as 1.38
$\mu _{B}$ has been found. The approach used can be called the quasi-atomic approach \cite{10,11} as the starting point
for the description of a solid, containing open-shell atoms, is the consideration of the atomic-like low-energy
electronic structure of the constituting atoms/ions, in the present case of the Co$^{2+}$ ions.

\section{Theoretical outline}

The used Hamiltonian is the same as we have used earlier for FeBr$_{2}$ and NiO \cite{10,11,12}. Our Hamiltonian for
CoO consists of two terms: the
single-ion-like term $H_{d}$ for the outer seven $d$ electrons and the $d$-$%
d $ intersite spin-dependent term, important for the formation of the magnetic state. We treat the 7 outer electrons of
the Co$^{2+}$ ion as forming the highly-correlated electron system 3$d^{7}$. The correlations among electrons in the
unfilled 3$d$ shell are approximated by two Hund's rules, that yield the ground-term quantum numbers $S$=3/2 and $L$=3,
i.e. the ground term $^{4}F$ \cite{13,14}. Such the localized highly-correlated electron system interacts in a solid
with the charge (it is known as crystal-field interactions) and spin surroundings. The charge surrounding has the
octahedral symmetry owing to the $NaCl$-type of structure of CoO. We take into account the small tetragonal distortion
as is experimentally
observed \cite{7}. For the quasi-atomic single-ion-like Hamiltonian of the 3$%
d^{7}$ system we take into account the crystal-field interactions of the octahedral symmetry, with the octahedral CEF
parameter $B_{4}$=-40 K like we
have assumed in our earliest paper \cite{15} (on basis of our recent studies %
\cite{16} a value of -30 K seems to be more appropriate, but it affects only very slightly the presented results as
$B_{4}$ determines excitations in the optical region; the most important is the negative sign that is determined by
negative charges of the oxygen octahedron), the spin-orbit interactions
with the spin-orbit coupling $\lambda _{s-o}$=-260 K like in the free ion %
\cite{13}, p. 399, and a small tetragonal distortion approximated by the term $B_{2}^{0}$=-10 K. The calculated
single-ion states under the octahedral crystal field and the spin-orbit coupling (the CoO$_{6}$ complex) are presented
in Fig. 1 of Ref. \cite{9}. All states have at least double Kramers degeneracy in agreement with the Kramers theorem.
Magnetic interactions are considered within the mean-field approximation with the molecular-field coefficient $n$ = -40
T/ $\mu _{B}$, that yields the experimental value T$_{N}$ of 290 K.

\section{Results and discussion}

For the purely octahedral symmetry the ground-state doublet is characterized by the total moment of $\pm $2.21 $\mu
_{B}$ ($\pm $ corresponds to the Kramers states). This moment is built up from the orbital and spin moments of $\pm
$0.55 and $\pm $1.66 $\mu _{B}$, respectively. The presence of the tetragonal distortion B$_{2}^{0}$=-10 K causes a
slight increase of these moments, to $\pm $2.54 $\mu _{B}$, $\pm $0.70 and $\pm $1.84 $\mu _{B}$, respectively. The
presence of the magnetic field, external or internal in case of the magnetically-ordered state polarizes two doublet
states. We have calculated that the Co moment in the magnetic state at 0 K experiences the molecular field of 161 T.
This field gradually diminishes to T$_{N}$ of 290 K. In the magnetic state there appears a spin-like gap of 40 meV (see
Fig. 2 of Ref. \cite{9}).

At 0 K in the magnetically-ordered state the magnetic moment was calculated to be 4.02 $\mu _{B}$. It is built up from
the spin moment of 2.64 $\mu _{B}$ ($S_{z}$=1.32) and the orbital moment of 1.38 $\mu _{B}$. The calculated total
moment is in good agrement with the very recent experimental datum of 3.98$\pm $0.06 $\mu _{B}$ for the Co moment
\cite{8}. The calculated orbital moment is quite substantial being more than 34\% of the total moment. Our theoretical
outcome, revealing the substantial orbital moment is in nice agreement with the recent magnetic x-ray experimental
result that has revealed the $L$/$S$ ratio of 0.95 \cite{17}- the calculated by us values lead to $L$/$S$ ratio, in
fact their $z$ components, of 1.04.

It is trivial to remind that the evaluation of the orbital moment is possible provided the spin-orbit coupling is taken
into account. Our calculations of the orbital moment of the 3$d$ atom and mostly the growing experimental evidence
\cite{17} for the existence of the significant orbital moment in the 3$d$-atom compounds confirm the importance of the
spin-orbit coupling for the description of the 3$d$ magnetism. The present model allows, apart of the ordered moment
and its spin and orbital components to calculate many physically important properties like temperature dependence of
the magnetic susceptibility, temperature dependence of the heat capacity, the spectroscopic $g$ factor, the fine
electronic structure in the energy window below 4 eV with at least 28 localized states. Although we consider here the
Co$^{2+}$ ion in the oxygen octahedron, the physical relevance of the obtained results to bulk CoO is obvious - the
$NaCl$ structure is built up by the edge sharing Co$^{2+}$ octahedra and the similar electronic structure occurs at
each Co site.

\section{Conclusions}

The orbital and spin moment of the Co$^{2+}$ ion in CoO has been calculated within the quasi-atomic approach taking
into account the intra-atomic
spin-orbit interactions by the same approach as we have earlier used for NiO %
\cite{12}, for FeBr$_{2}$ \cite{11} and for LaCoO$_{3}$ \cite{16}. The orbital moment of 1.38 $\mu _{B}$ amounts at 0
K, in the
magnetically-ordered state, to more than 34\% of the total moment of 4.02 $%
\mu _{B}$. In our description we largely employ the atomic physics. We use only four parameters, B$_{4}$, $\lambda
_{s-o}$, B$_{2}^{0}$ and $n$ - all of them have the clear physical meaning. Our atomic-like approach can be extended to
take into account other physical phenomena. Our approach takes into account very strong correlations among $d$
electrons in agreement with the general conviction about their fundamental importance in 3$d$ oxides. We take these
correlations to be stronger than crystal-field interactions. In contrary to Ref. \cite{8}, where an effective charge of
+1.66 only has been found at the Co site we work with the +2.0 charge. According to our calculation the +2.0 state
lowers the cohession energy by 45 \% - so our solution is energetically favoured. Our theoretical model can be regarded
as
a model with $U$ = $\infty $ and the Hund coupling constant $J_{H}$ = $%
\infty $. Our model, presenting here in the purely ionic form, explains the insulating ground state despite the open
3$d$ shell. Our studies indicate that it is the highest time in solid-state physics to ''unquench'' the orbital moment
in the description of 3$d$-atom containing compounds.\newline

$^\ast$ This version has been resubmitted 8 October 2003 to Phys. Rev. B as LS8314. It is the revised version of the
original submission 14-05-2002 to Phys.Rev.Lett. as Orbital moment in CoO. Later, 27-03-2003, it was transferred to
Phys.Rev.B that is available at cond-mat/0211705.

\newpage

\begin{center}
\textbf{APPENDIX A}\\[0pt]
\textbf{{\small {Submission letter to Physical Review Letters of 14-05-2002}}%
}
\end{center}
From: \quad sfradwan@cyf-kr.edu.pl \quad To: \quad prltex@aps.org\newline Subject: \quad submit PRL Radwanski \quad
Date sent: \quad Tue, 14 May 2002 21:03:12 +0200\newline From R.J. Radwanski and Z. Ropka 14.05.2002 \quad To: \quad
Editor of Phys.Rev.Lett.

Dear Sir,

Please find attached my paper: The orbital moment in CoO by R.J. Radwanski, Z.Ropka, that I submit for publication in
Phys.Rev. Lett..

Paper is prepared in the tex.file. Paper contains three Figures that are attached in EPS.file.The paper offers novel
theoretical result for the orbital and spin moment in CoO. Such calculations have not been published up to now. It is
the first consistent calculations for the orbital moment. Our calculations are taking into account the spin-orbit
coupling. Practically all calculations presented up to now in literature were not able to calculate orbital moment. The
orbital magnetism is a hot topic in the nowadays magnetism.

The paper accounts 4 pages, Plus figure caption (page 5) and 3 figures in eps.

We would appreciate publication of our paper.

Sincerely Yours, R.J.Radwanski\\
Attachments: C:$\backslash$3d-02$\backslash$CoO-prl$\backslash$Radwanski-CoO.tex,

Fig1-Radw.eps, Fig2-Radw.eps, Fig3-Radw.eps\\

\bigskip
\begin{center}
\textbf{APPENDIX B}\\[0pt]
\textbf{{\small {Rejection by the Editor of Physical Review Letters of 24-05-2002}}}
\end{center}
\bigskip
Due to reasons independent on authors it is only available on www.css-physics.edu.pl
\bigskip
\begin{center}
\newpage
\textbf{APPENDIX C}\\[0pt]
\textbf{{\small {Resubmission letter to Physical Review Letters of 27-05-2002%
}}}
\end{center}
From: \quad sfradwan@cyf-kr.edu.pl \qquad R.J. Radwanski and Z. Ropka\newline To: \quad Physical Review Letters
$<$prl@ridge.aps.org$>$ \newline Subject: \quad resub LS8314, 3 fig. like before \quad Date sent: Mon, 27 May 2002
22:19:21 +0200

To Editor of Phys.Rev.Lett. Reinhardt B. Schuhmann

concerns: LS8314 -The orbital moment in CoO

Dear Editor,

In the answer to your fast rejection of our a/m Letter, please take at first the sincere congratulations for becoming
the Editor of \ Phys.Rev.Lett.. It is, however, very responsible job. Secondly, great thanks for fast answer. However,
please be so kind to say a word why our Letter is not suitable for the publication in Phys.Rev.Lett. Our Letter
presents, for the first time according to our knowledge, the calculations of the orbital moment in CoO (if not, please
write where it was published). The topic of the orbital moment is very important in 3d magnetism, see just published
papers in Phys.Rev. B and J.Phys.Chem. Solids mentioned in our reference list.

We resubmit our Letter - 3 figures are as previously. Hoping to hear from you good news

Sincerely Yours, R.J. Radwanski and Z. Ropka\\
Attachments: \ C:$\backslash$3d-02$\backslash$CoO-prl$\backslash$Radwanski-CoO.tex\\

\begin{center}
\textbf{APPENDIX D}\\[0pt]
\textbf{{\small {Secend rejection by the Editor of Physical Review Letters of 05-June-2002}}}
\end{center}
\bigskip Due to reasons independent on authors it is only available on www.css-physics.edu.pl
\bigskip
\newpage
\begin{center}
\textbf{APPENDIX E}\\[0pt]
\textbf{{\small {Submission to Phys. Rev. B 27.03.2003 - transfer from Phys. Rev. Lett.}}}
\end{center}
From: \quad sfradwan@cyf-kr.edu.pl\newline To: \quad prbtex@aps.org\newline Subject: \quad submit PRB Radwanski\newline
Date sent: \quad Thu, 27 Mar 2003 20:50:09 +0100\newline From: \quad R.J. Radwanski and Z. Ropka 27.03.2003\newline
To:\quad Editor of Phys.Rev. B\newline

Dear Editor,

Dear Dr P.D. Adams

Please find attached our paper: The orbital moment in CoO by R.J. Radwanski, Z.Ropka, that I submit for publication in
Phys.Rev. B.

Paper is prepared in the tex.file. Paper accounts 4 pages, 16 Refs, three Figures that are attached in EPS.file.

The paper offers novel theoretical result for the orbital and spin moment in CoO. Such calculations have not been
published up to now. It is the first consistent calculations for the orbital moment. Our calculations are taking into
account the intra-atomic spin-orbit coupling. Practically all calculations presented up to now in literature were not
able to calculate orbital moment.

The orbital magnetism is a hot topic in the nowadays magnetism. The direct motivation for this paper was a publication
by Jauch and Reehuis in Phys.Rev. B 65 (2002) 125111 as well as earlier PRB 64 (01) 052102, PRB 44 (91) 943 and 1364,
PRB 49 (94) 10864 and 10170; Phys.Rev.Lett. 80 (98) 5758.

We would appreciate the publication of our paper and the scientific referee process.

Sincerely Yours,

R.J.Radwanski\newline
Attachments: C:$\backslash$3d-02$\backslash$CoO-prl%
$\backslash$Radwanski-CoOb.tex, \newline \qquad\qquad Fig1-Radw.eps, Fig2-Radw.eps, Fig3-Radw.eps
\bigskip

\begin{center}
\textbf{APPENDIX F}\\[0pt]
\textbf{{\small {Rejection by the Editor of Physical Review B of 15-09-2003}}%
}
\end{center}
\bigskip Due to reasons independent on authors it is only available on www.css-physics.edu.pl
\newpage
\begin{center}
\textbf{APPENDIX G}\\[0pt]
\textbf{{\small {RESUBMISSION\ TO\ Phys. Rev.\ B\ 8 October 2003}}}
\end{center}

From: \quad sfradwan@cyf-kr.edu.pl\newline To: \quad Physical Review B $<$prbtex@ridge.aps.org$>$
\newline
Subject: \quad resub PRB LS8314B Radwanski\newline Date sent: \quad Wed, 8 Oct 2003 18:04:26 +0200\newline From: \quad
R.J. Radwanski Krakow, 08-10-2003\newline Re:\quad LS8314B: $>$ Orbital moment in CoO, By: R.J. Radwanski and Z. Ropka

submitted 14-05-02 to PRL, transferred to PRB 28-03-03.

Dear Editor of Phys. Rev. B,

In the answer to your last e-mail of 15-09-03 of Julie Kim-Zajonz, about our apper LS8314B: Orbital moment in CoO, we
say that we highly appreciate great physical knowledge of the Editor of PRB, however, than the Editor should openly
publish his scientific view, as extremaly scientifically important as conventional paper. Otherwise, a nonsense
situation is created that very important scientific problems are discussed in private letters whereas in Phys. Rev. are
published papers on much lower importance.

I hope that we agree that this situation is highly improper. Please understand that in this way the Editor of Phys.Rev.
B, being aware of this fact or not, manipulates Physics.Thus we propose to forget about it and we ask the Editor of PRB
to follow the scientific procedure - we remind we chose PRB due to just published theoretical paper about CoO (PRB 65
(02) 125111).

Two statements, given as example in your last e-mail and added to a standard rejection letter of Editor of Phys. Rev. B
are wrong:

1. ''the manuscript does not appear to contain enough information for a reader to reproduce the results'' - we stress
that in our paper there are all (exactly all) information that a physicist who knows the standard book of Van Vleck
from 1932 can reproduce our results (but please do not use an argument that our approach is too old),

2. ''the manuscript ... also lacks concrete scientific arguments as to why your approach is superior.'' - At first, we
calculate the orbital moment, 1.38 miu-bi, whereas the paper PRB 65, 125111 does not calculate this physical value. In
the last paper the orbital moment is inferred as 1.6 miu-bi (p. 7, left column, line 3 bottom) obtained by the simple
substraction of the calculated spin-moment, of 2.40 miu-bi, p. 7, left, 23top) from the know-from-literature total
moment of 3.98 (p. 7, left, line 4 b). Is it enough for your demand for a proof of the superiority? Please remember,
that up to now the orbital moment was not calculated -according to the paper PRB 65, 125111 there were only rough
prediction for the orbital moment in size of 1 miu-bi as is mentioned on p. 7 right column, l 2 t after two Refs 23,26.
Ref. 23 it is PRL paper from 1998.

For us our approach is very natural. It is in agreement with all modern solid-state physics understanding like
atomistic building of matter, atomic physics, strong correlations, local symmetry, spin-orbit coupling, ... Your
objection that our paper ''lacks concrete scientific arguments for the superiority of our approach'' would indicate
that Editor of Phys.Rev. B does not accept this understandings and prefers not well defined poetrical concepts. We
take, however, these two sentences as a confirmation by the Editor PRB of the originality and the novelty of our
approach. In such the situation, being fully convinced about importance of our paper and knowing that the paper on this
problem just appears in Phys. Rev. B, 65 (2002) 125111, we resubmit our paper in a shorter form (in order to save the
space in the journal) without figures. In order to satisfied the remarks of the Editor of PRB i) we stressed that all
four parameters needed for the reproduction of our results are in our paper and ii) we added some scientific comments
and arguments for the superiority of our approach.

Paper contains 4 pages + 17 references, mainly from Phys.Rev.. All 3 figures are skipped.

Owing to the publication of the paper 65 (2002) 125111 in Phys. Rev. B, the submision of our paper about CoO to Phys.
Rev. B is fully justified. We would appreciate the scientific procedure of our paper and its fast publication in
Phys.Rev. B. It enables the open scientific discussion.

Sincerely Yours,

R.J. Radwanski and Z. Ropka\\
Attachments: C:$\backslash$3d-02$\backslash$CoO-prl$\backslash$Radwanski-CoO-X-03.tex
\end{document}